\documentclass{cimento}
\usepackage{graphicx}  
\usepackage{amsmath,amssymb,amsfonts}

\title{The power of photons: Cavity-mediated energy transfer between quantum devices}
\author{A.~Crescente\from{ins:x}\from{ins:y}}
\instlist{\inst{ins:x} Dipartimento di Fisica, Universit\`a di Genova, Via Dodecaneso 33, 16146, Genova, Italy
  \inst{ins:y} CNR-SPIN, Via Dodecaneso 33, 16146, Genova, Italy}

\begin{document}

\maketitle

\begin{abstract}

The coherent energy transfer between a quantum charger and a quantum battery is analyzed. In particular, we study how to improve the direct energy transfer by adding a photonic cavity as a mediator.
We show that the additional degree of freedom given by the photons consistently improves the transfer performances, above all in the off-resonant case, where there is a mismatch in the energy levels. 
An experimental feasible way to switch-on and off the interaction between each part of the systems and the possibility of changing the energy levels mismatch will be described, in view of finding the best working setup. 

\end{abstract}

\section{Introduction}

In the last decades quantum technologies have assumed a central role in the scientific research worldwide~\cite{Riedel17, Raymer19}. In this framework particular interest has been devoted to the new and fast growing field of quantum batteries (QBs)~\cite{Alicki13, Binder15, Campaioli17}.
Most of the research on QBs has been focussed on finding efficient ways to store energy into a quantum system and release it on demand~\cite{Ferraro18, Andolina18, Crescente20, Crescente20b, Santos21, Delmonte21, Gemme22, Benenti22, Shaghaghi22, Erdman22}, in order to locally supply energy to miniaturized devices. However, only few works addressed the interesting and still largely unexplored problem of coherent energy transfer between distant quantum systems~\cite{Andolina18, Farina19}. In this direction, the realization of energy transfer processes in the quantum domain could represent a crucial step towards the creation of a capillary energy network able to connect distant parts of a fully quantum device with improved performances~\cite{Scarlino19}.

The aim of this work is to characterize the coherent energy transfer between two quantum systems, focussing on the simple, but experimentally relevant~\cite{Sillanpaa07, Dicarlo09}, situation of two two-level systems (TLSs)~\cite{Weiss, Paladino08, Sassetti96}, the first playing the role of a quantum charger and the second being the QB.
In the following, we compare the already well known direct energy transfer process with a cavity-mediated one, where photons act as a quantum bus for the energy transfer.
We investigate the stored energy in the different parts of the total system both on-resonance, i.e. when the level spacing of the two TLSs and the frequency of the photons in the cavity are the same, and off-resonance, namely when there is a mismatch in the TLSs level spacing and photons frequency. This latter analysis is justified by the fact that experimentally it is difficult to realize absolutely identical TLSs~\cite{Scarlino19, Sillanpaa07}. The main result of this work is the demonstration that the presence of the cavity, as a mediator, allows a faster energy transfer between the charger and the QB, compared to the direct model. In addition, the photons in the cavity allows improving the energy transfer when the system is off-resonance.
A possible experimental feasible way to switch on and off the interaction and the possibility of changing the mismatch in the TLSs level spacing will be considered in order to find the best performances of the device.

This paper is organized as follows. In Section~\ref{Model} we introduce the cavity-mediated model and we briefly recall the direct coupling scenario. In Section~\ref{FM} the usual figures of merits are considered with particular emphasis on the transferred energy and work done in switching on and off the interaction. Section~\ref{Results} enlightens the results obtained in the cavity-mediated model, showing its better performances compared to the direct one. In Section~\ref{Mismatch} it is shown how it is possible to control the mismatch between the TLSs, to emulate a process in which the interaction is switched on and off. Finally Section~\ref{Conclusion} is devoted to the conclusions.

\section{Cavity-mediated energy transfer model}\label{Model}

\begin{figure}[h!]
\centering
\includegraphics[scale=0.17]{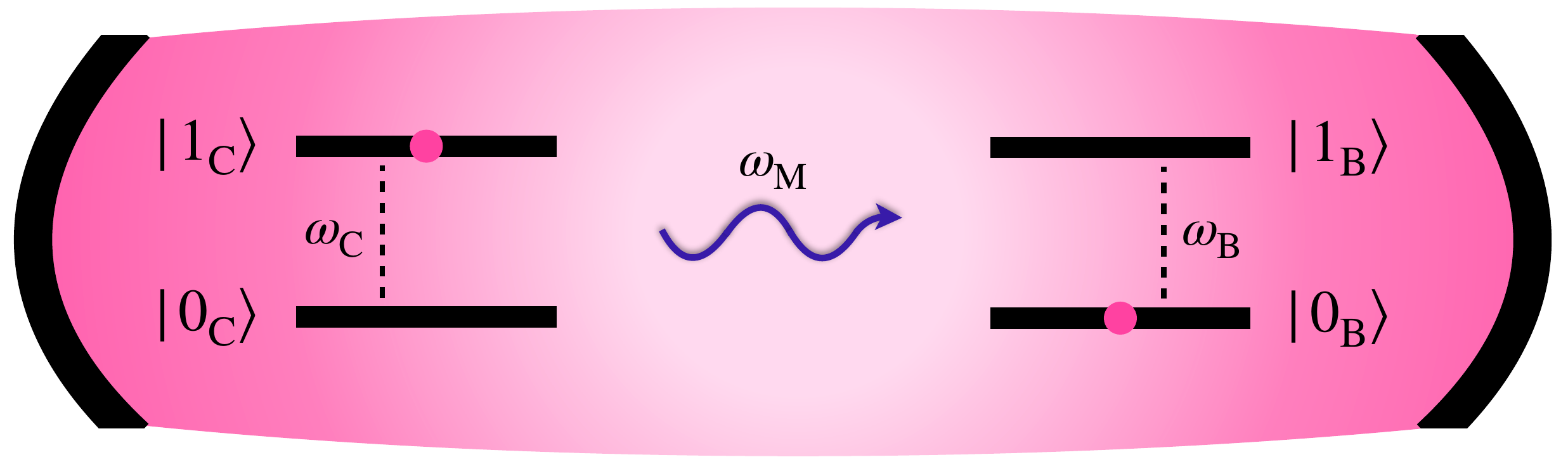} 
   \caption {Schematic representation of the cavity-mediated process where photons of frequency 
   $\omega_{\rm M}$ mediate the energy transfer between the charger and the QB, modeled as TLSs, with energy separation $\omega_{\rm C}$ and $\omega_{\rm B}$ respectively.}
   \label{fig1}
\end{figure}

The setup for the energy transfer between a quantum charger (C) and a QB (B) in presence of a cavity (M) acting as a mediator is shown in Figure~\ref{fig1}. In order to keep the analysis relevant for possible experimental implementations, both the charger and the QB are modeled as TLSs with ground state $|0_{\rm C,B}\rangle$ and excited state $|1_{\rm C,B}\rangle$ respectively. Then the free Hamiltonian of the system can be written as
\begin{equation}\label{H0} H_0=H_{\rm C}+H_{\rm B}+H_{\rm M}=\frac{\omega_{\rm C}}{2}\sigma_z^{\rm C}+\frac{\omega_{\rm B}}{2}\sigma_z^{\rm B}+\omega_{\rm M}a^\dagger a, \end{equation}
where $\omega_{\rm C}$ and $\omega_{\rm B}$ are the energy separation of the charger and QB respectively and $\omega_{\rm M}$ is the frequency of the photons in the cavity. Here, $\sigma_z^i$ is the Pauli matrix along the $\hat z$ direction acting on the $i=\rm{C,B}$ space and $a$ ($a^\dagger$) is the annihilation (creation) operator of the photons.

The interaction Hamiltonian of the cavity-mediated model is represented experimentally by two superconducting qubit (C and B) interacting with a LC resonator (M)~\cite{Krantz19} and can be written as
\begin{equation}\label{Hintm} H_{\rm int,m}^{(t)}= gf(t)[a^\dagger(\sigma_-^{\rm C}+\sigma_-^{\rm B})+a(\sigma_+^{\rm C}+\sigma_+^{\rm B})], \end{equation}
where to simplify the problem we have assumed the same coupling constant $g$ both between the mediator and the charger and between the mediator and the QB. Here, the apex $(t)$ indicates the parametric dependence of the Hamiltonian on time and $\sigma_\pm=(\sigma_x\pm \sigma_y)/2$ are the spin ladder operators, with $\sigma_{x,y}$ the Pauli matrix along the $\hat x, \hat y$ directions. Moreover, $f(t)$ is a dimensionless time dependent function which has been introduced in order to take into account the switching on and off of the interaction. Its precise shape will be specified later.

In order to simplify the solution of the dynamics, we have considered the Hamiltonian in Eq.~(\ref{Hintm}) in the rotating-wave approximation (RWA)~\cite{Schweber67, Graham84, Schleich}, where a constraint on admissible values for the coupling constant $g \lesssim 0.1\omega_{\rm C, B}$ is imposed. 

Notice that this approximation does not represent a major limitation due to the fact that most of the experimental realizations of such quantum systems well fits into this regime~\cite{Majer05, Niskanen07}.

The complete Hamiltonian for the cavity-mediated model is then given by
\begin{equation}\label{HM} H_{\rm m}^{(t)}=H_{\rm C}+H_{\rm B}+H_{\rm M}+H_{\rm int,m}^{(t)}. \end{equation}
In the following we will investigate both the resonant $\omega_{\rm C}=\omega_{\rm M}=\omega_{\rm B}$ and off-resonant regime $\omega_{\rm C}=\omega_{\rm M}=\alpha\omega_{\rm B}$, with $\alpha$ a positive real parameter.
Notice that it is important to consider off-resonance conditions since experimentally it is difficult
to realize identical TLS in solid state platforms~\cite{Scarlino19, Sillanpaa07}.

\subsection{A reference model: the direct coupling}

Here we briefly recall the direct coupling model between the charger and the QB, already analyzed in Ref.~\cite{Andolina18, Crescente22}, that will be considered in the following as a comparison with the cavity-mediated case. Under the assumption of a local (short range) and direct capacitive coupling between the TLSs the interaction Hamiltonian in the RWA assumes the following form~\cite{Schweber67, Graham84, Schleich}
\begin{equation} H_{\rm int,d}^{(t)}= gf(t)(\sigma_-^{\rm C}\sigma_+
^{\rm B}+\sigma_+^{\rm C}\sigma_-^{\rm B}), \end{equation}
where $g$ is the coupling constant between the TLSs and $f(t)$ is the same time dependent function introduced before. Consequently the complete Hamiltonian for the direct energy transfer is
\begin{equation} H_{\rm d}^{(t)}= H_{\rm C}+H_{\rm B}+H_{\rm int,d}^{(t)}. \end{equation}

\section{Figures of merit}\label{FM}

The main task of this Section is to characterize the energy transfer process between the charger and the QB, taking also into account the switching on and off of the interactions.

\subsection{Stored energy and transfer time}

The stored energy at time $t$ in $i=\rm{C, B, M}$ is given by
\begin{equation}\label{Et} E_{ i}(t)\equiv{\rm Tr}\{\rho(t) H_{i}\}-{\rm Tr}\{\rho(0) H_{i}\}, 
\end{equation}
where ${\rm Tr}\{\dots \}$ represents the conventional trace operation, $\rho(0)=|\psi(0)\rangle \langle \psi(0)|$ is the total density matrix of the system at the initial time $t=0$ and $\rho(t)$ is the time evolved density matrix according to the Hamiltonian in Eq.~(\ref{HM}). In the following we are interested in studying the energy transfer between a full charger, whose initial state is $|1_{\rm C}\rangle$, and an empty QB, with initial state $|0_{\rm B}\rangle$. For the cavity we will consider as initial state Fock states with $n$ photons, namely $|n\rangle$. Consequently the total initial state assumes the form
\begin{equation}\label{IS} |\psi(0)\rangle=|1_{\rm C},0_{\rm B},n\rangle. \end{equation}

In analogy to what done in Eq.~(\ref{Et}) it is also useful to consider the energy associated to the interaction term, defined as 
\begin{equation}\label{Eint} E_{\rm int}(t)\equiv{\rm Tr}\{\rho(t) H_{\rm int, m}^{(t)}\}-{\rm Tr}\{\rho(0) H_{\rm int, m}^{(0)}\}. \end{equation}
Notice that, due the chosen initial state in Eq.~(\ref{IS}) and the form of $H_{\rm int,m}^{(t)}$ in Eq.~(\ref{Hintm}), the condition
${\rm Tr}\{\rho(0) H_{\rm int, m}^{(0)}\}=0$ is always verified,
which further simplifies Eq.~(\ref{Eint}).
Moreover, we define 
\begin{equation} 
\label{Emax} E_{\rm B, \rm max} \equiv E_{\rm B}(t_{\rm B, \rm max}),
\end{equation}
namely the first local maximum achievable value of the stored energy in the QB, which occurs at the shorter charging time $t_{\rm B, \rm max}$ and with 
\begin{equation} 
\label{Emax} \bar E_{\rm C} \equiv E_{\rm C}(t_{\rm B, \rm max}),
\end{equation}
the value of the energy in the charger at the same time. Indeed, as we will show below, while at resonance all the maxima are obtained at the same times, out of resonance this could not be the case.

Similar considerations on the stored energy and initial state of the system can be done for the direct coupling case, see Refs.~\cite{Andolina18, Crescente22}.

\subsection{Average work done to switch on and off the interaction}

To fully characterize energy transfer processes it is necessary to consider also the power employed for the switch on and off of the interaction.
Here, we recall its formal definition
\begin{equation} P(t)\equiv \frac{d}{dt}[{\rm Tr}\{\rho(t)H^{(t)}_i\}]={\rm Tr}\bigg\{\rho(t) \frac{\partial H_{\textrm{int},i}^{(t)}}{\partial t}\bigg\},\end{equation}
where $i=\rm{m, d}$ for the cavity-mediated model and for the direct one, respectively.
The corresponding average work $W(t)$ at a given time $t$ is then given by
\begin{equation} 
W(t)=\int_0^t dt'P(t').
\label{W_t_gen}
\end{equation}
Specifying to the cases of cavity-mediated and direct coupling the powers can be written as
\begin{eqnarray} 
P_{\rm m}(t)&=&\frac{dE_{\rm C}(t)}{dt}+\frac{d E_{\rm B}(t)}{dt}+\frac{d E_{\rm M}(t)}{dt}+\frac{d E_{\rm int,m}(t)}{dt}  \\
P_{\rm d}(t)&=&\frac{dE_{\rm C}(t)}{dt}+\frac{d E_{\rm B}(t)}{dt}+\frac{d E_{\rm int,d}(t)}{dt}.
 \end{eqnarray}

\noindent The corresponding works obtained from. Eq.~(\ref{W_t_gen}) fulfill the following conservation energy relations
\begin{eqnarray}\label{CEdirect}
W_{\rm m}(t)&=&E_{\rm C}(t)+E_{\rm B}(t)+E_{\rm M}(t)+E_{\rm int,m}(t) \\
\label{CEmed}
W_{\rm d}(t)&=&E_{\rm C}(t)+E_{\rm B}(t)+E_{\rm int,d}(t). 
\end{eqnarray}

In both cases, reintroducing the off-resonance parameter $\alpha$, it is possible to obtain simplified forms for the work~\cite{Crescente22}, leading to a common equation for both models
\begin{equation}
\label{Work} 
W(t)=(1-\alpha)E_{\rm B}(t)+E_{\rm int,i}(t), 
\end{equation}
with $i=\rm{m, d}$.

\subsection{Functional form of the switch on and off function}

\begin{figure}[h!]
\centering
\includegraphics[scale=0.42]{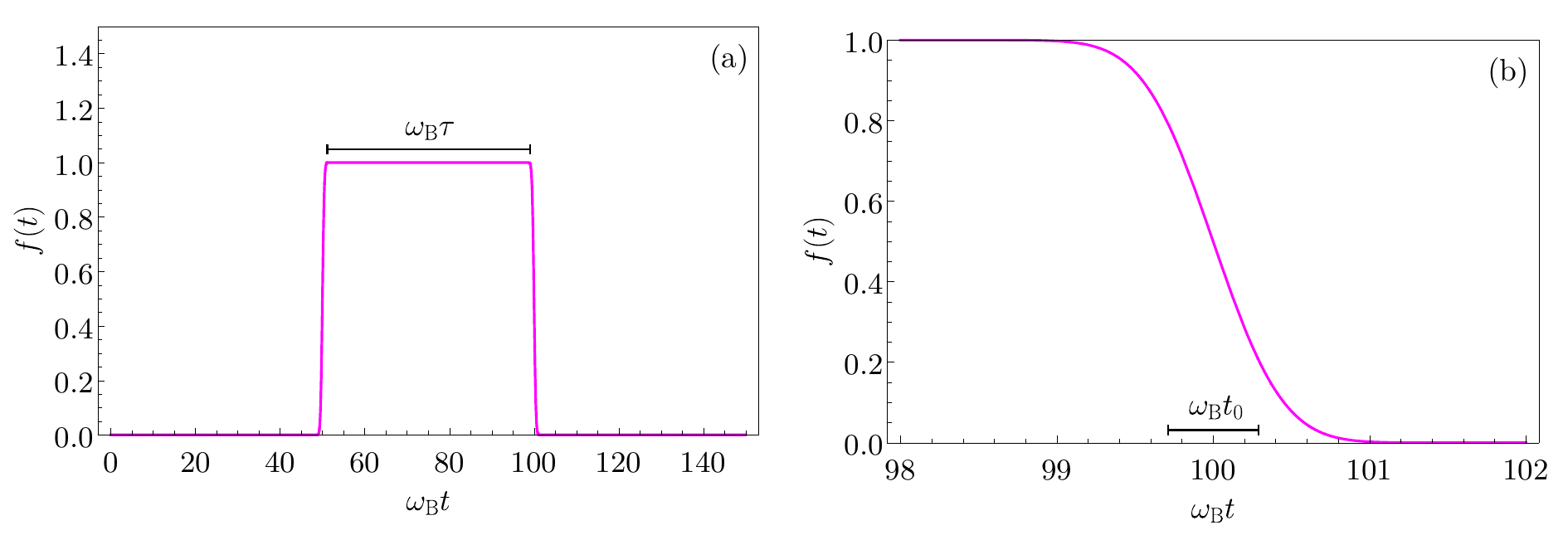} 
   \caption {Panel (a): Behaviour of $f(t)$ as function of $\omega_{\rm B}t$ for $\omega_{\rm B}\tau=50$ and $\omega_{\rm B}t_0=0.1$. Panel (b): Zoom of the same function in correspondence of the switching off region.}
   \label{fig3}
\end{figure}

Here, the form of the switching on and off function $f(t)$ in Eq.~(\ref{HM}) is specified. From now on, the following functional form is considered (see Figure~\ref{fig3}) 
\begin{equation} 
f(t)= \frac{\mathrm{erf}\bigg(\dfrac{t-\tau}{t_0}\bigg)-\mathrm{erf}\bigg(\dfrac{t-2\tau}{t_0}\bigg)}{2\mathrm{erf}\bigg(\dfrac{\tau}{2t_0}\bigg)},
\label{f_t}
\end{equation}

\noindent which describes a smooth switching on and off of the interaction between the charger and the mediator and the mediator and the QB in the cavity-mediated model or between the two quantum systems, charger and QB, in the direct one.

From Figure~\ref{fig3} it can be seen that the parameter $\tau$ controls the time window where the interaction is active, while $t_0$ is the width of the switching ramp. 
By controlling the parameters $\tau$ and $t_0$ it is then possible to turn off the interaction when the first maximum of the energy stored in B is achieved, meaning that at time $t_{\rm B, max}$ the interaction Hamiltonian $H_{\rm int}^{(t)}$ is switched off.

\newpage

\section{Results}\label{Results}

In this Section we report and discuss the main results for the cavity-mediated energy transfer and compare them with the direct model. All the results are obtain through an exact numerical diagonalization (see Ref.~\cite{Crescente22} for more details).

Notice that in the following we will consider the composite system as a closed quantum system, meaning that dissipative effects related to relaxation and dephasing phenomena are not taken into account. This is possible when the typical relaxation $t_r$ and dephasing $t_\varphi$ times are longer with respect to the considered evolution time $t$, i.e. $t_r, t_\varphi \gg t$~\cite{Sassetti96, Devoret13, Wendin17, Carrega20}.

\subsection{Enhancing the energy transfer performances}

\begin{figure}[h!]
\centering
\includegraphics[scale=0.4]{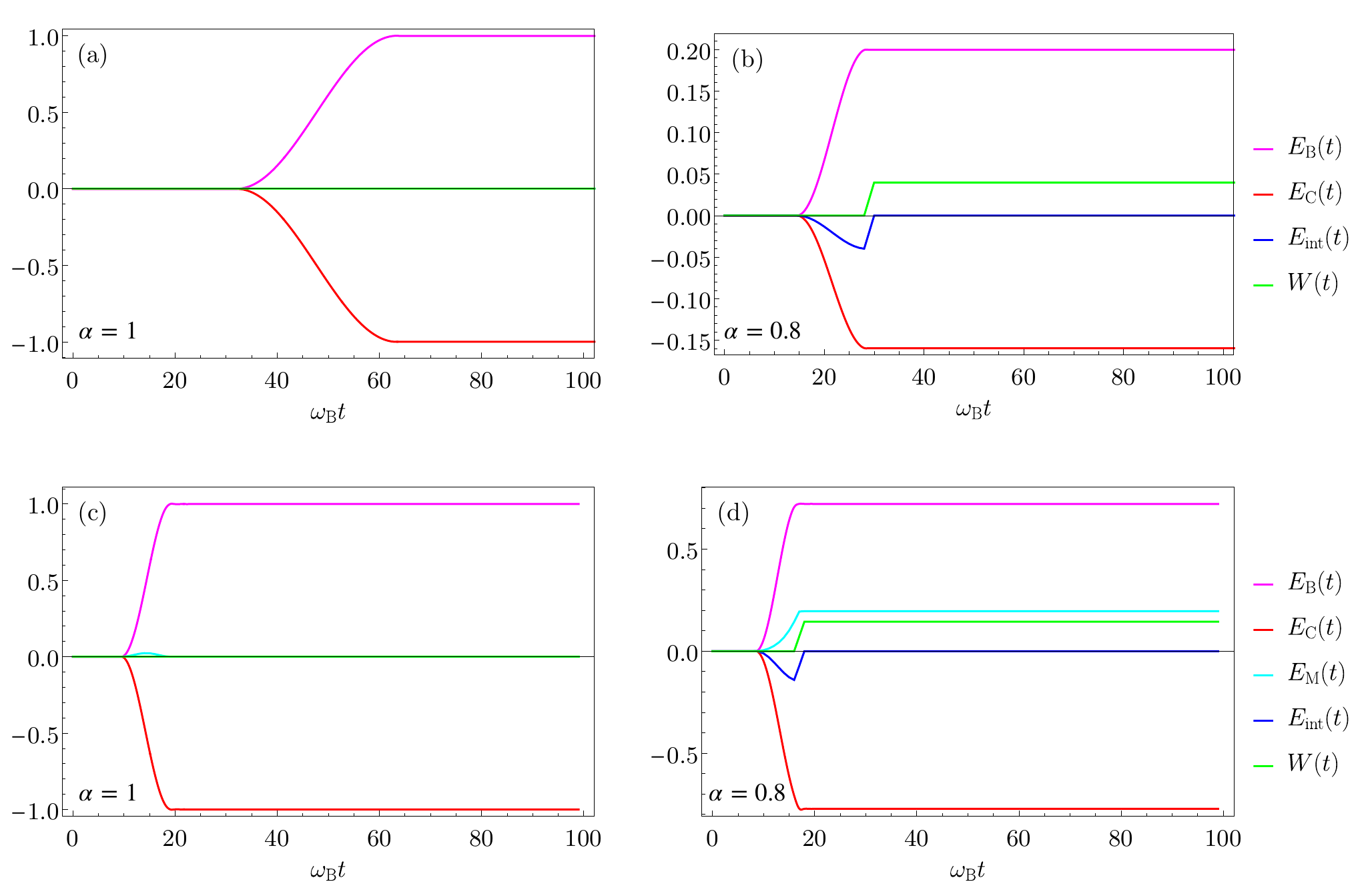} 
   \caption {Behaviour in units of $\omega_{\rm B}$ of $E_{\rm B}(t)$ (magenta curves), 
   $E_{\rm C}(t)$ (red curves), $E_{\rm M}(t)$ when present (cyan curves), $E_{\rm int}(t)$ (blue curves) and $W(t)$ (green curves). In panel (a) and (b) we report the direct coupling as a reference, while panel (c) and (d) represent the cavity-mediated scenario. The curves are considered in the resonant regime $\omega_{\rm B}=\omega_{\rm C}$ in panel (a) and (c) and in the off-resonant regime $\omega_{\rm C}=0.8\omega_{\rm B}$ in panel (b) and (d). The values $\omega_{\rm B}\tau=32$ (a), $\omega_{\rm B}\tau=14.5$ (b), $\omega_{\rm B}\tau=9.5$ (c) and $\omega_{\rm B}\tau=8.5$ (d) are considered respectively yo switch off the interaction when the first maximum of the transferred energy is achieved. Other parameters are $g=0.05\omega_{\rm B}$, $\omega_{\rm B}t_0=0.1$ and $n=10$.}
   \label{fig4}
\end{figure}

In Figure~\ref{fig4} the behaviour in time of the different terms involved in Eqs.~(\ref{CEdirect}) and~(\ref{CEmed}) is shown. The direct energy transfer between the charger and the QB is reported in panels (a) and (b), in order to compare it to the cavity-mediated case, in panels (c) and (d). The analysis is brought up both on-resonance ($\alpha=1$) and off-resonance for the representative value $\alpha=0.8$ for a given coupling constant $g=0.05\omega_{\rm B}$ and a number of photons $n=10$ in the cavity. 

\newpage

From Figure~\ref{fig4} (a) we observe that, when the system is on-resonance, the energy $E_{\rm C}(t)$ stored into the charger completely goes into the QB. Moreover, during this process the energy associated to the interaction $E_{\rm int}(t)$ and the work $W(t)$ remains zero, as expected from Eq.~(\ref{Work}). 
Different is the situation when the system is off-resonance [see Figure~\ref{fig4} (b)]. Here, the charger loses only a fraction of its energy $\sim 15\%$ and transfers it to the QB. In this case, another contribution to the energy transfer process is given by the interaction term, that together with the charger, allows reaching $\sim 20\%$ of the full charge in the QB.
Notice that off-resonance a finite amount of work is done in the transfer process, confirming the overall conservation discussed in Eq.~(\ref{CEdirect}).

For what concerns the cavity-mediated case, it is possible to exploit the additional degree of freedom offered by the number of photons $n$ to improve the performances of the energy transfer in the composite system.
When we consider a higher number of photons, e.g. $n=10$ in panel (c), we can observe that on-resonance it is possible to obtain a complete and faster energy transfer process compared to the one obtained in the direct case. More relevant is the impact of the richer structure of the mediator in the off-resonant case [see panel (d)]. Here the charger releases almost all its energy. Even if a fraction of this energy remains trapped into the mediator, it is possible to charge the QB more then $\sim 71 \%$ in a very short time. In this case the mediator plays the role of facilitator for the energy transfer, leading to a major improvement with respect to the direct case and giving an important impact for practical applications.

\subsection{The advantage of using a large number of photons}

\begin{figure}[h!]
\centering
\includegraphics[scale=0.43]{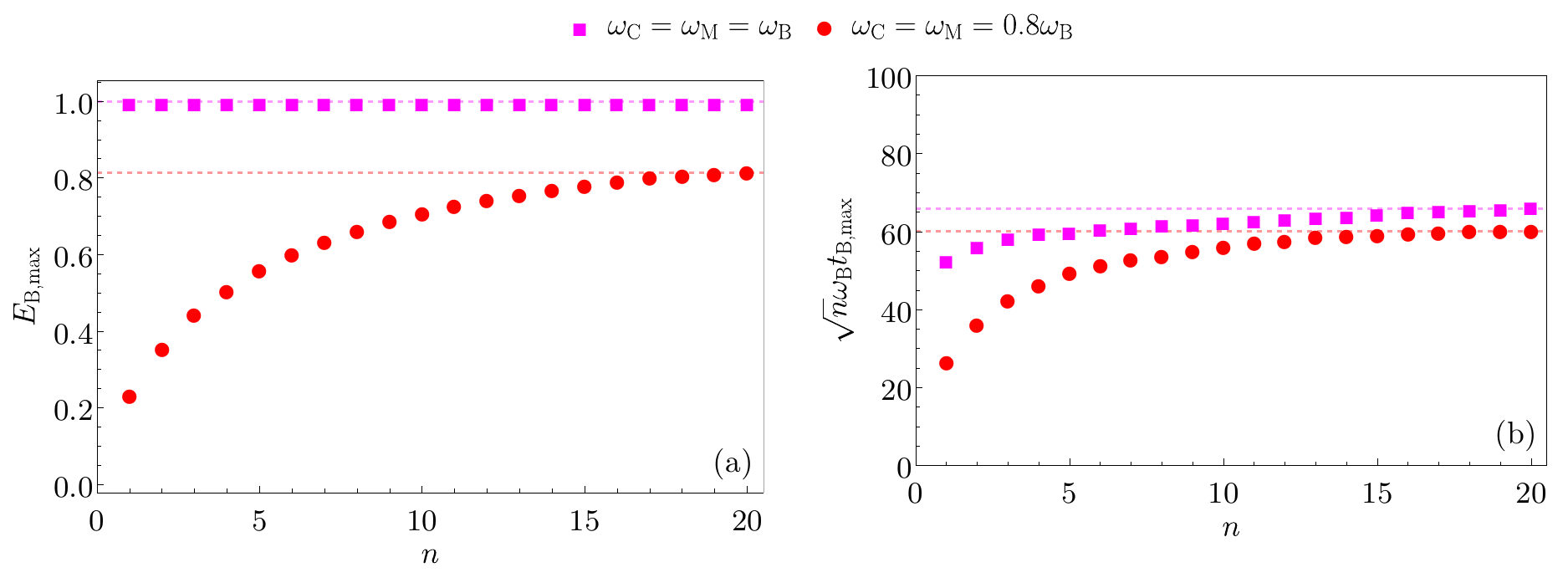} 
   \caption {Maximum of the stored energy in the QB $E_{\rm B, max}$ in units of $\omega_{\rm B}$ (a) and corresponding rescaled charging time $\sqrt{n}\omega_{\rm B}t_{\rm B,max}$ (b) as function of the number of photons $n$ for the cavity-mediated model for $\omega_{\rm C}=\omega_{\rm M}=\omega_{\rm B}$ (magenta full squares) and $\omega_{\rm C}=\omega_{\rm M}=0.8\omega_{\rm B}$ (red full dots). Other parameters are $g=0.05\omega_{\rm B}$ and $\omega_{\rm B}t_0=0.1$. Values of $\tau$ are chosen for each point in such a way to switch off the energy transfer in the system when the first maximum of the energy stored in the QB is achieved.}
   \label{fig5}
\end{figure}

The advantage of using a cavity as a mediator for the energy transfer is further enhanced by increasing the number of photons $n$ into the cavity. This can be seen from Figure~\ref{fig5}, where the maximum of the stored energy in the QB [panel (a)] and the energy transfer times [panel (b)] are reported as a function of $n$.
From panel (a) we can see that by increasing the number of photons in the cavity it is possible to consistently improve the energy transferred to the QB also in the off-resonant case. 

\newpage

In fact, for $\alpha=0.8$ at large $n$ we obtain a charging of the QB exceeding $\sim 80 \%$, which is even better than the one reported for $n=10$ in Figure~\ref{fig4} (d).
In principle, even if experimentally infeasible, it is possible to reach, even off-resonance, the maximum charge of the QB for large $n$, with $E_{\rm B, max}(n\rightarrow +\infty)\rightarrow \omega_{\rm B}$ (not shown).

The advantages in using a larger number of photons can also be seen from the charging times. Indeed, at large values of $n$ the energy transfer time scales as $t_{\rm{B, max}}\propto n^{-1/2}$ both on- and off-resonance [see asymptotes in Figure~\ref{fig5} (b)].

\section{Controlling the two-level systems mismatch}\label{Mismatch}
To conclude we discuss a protocol that allows controlling the TLSs mismatch, smoothly changing the energy separation of the TLSs. This is a way to mimic the switch on and off of the interaction introduced in Section~\ref{Model}, since when the TLSs are far off-resonance the energy transfer is greatly suppressed, while when they are on-resonance it is generally promoted.
Moreover, this kind of protocol is well-controlled in experiment concerning qubits~\cite{Kafri17, Arute19}. 

\begin{figure}[h!]
\centering
\includegraphics[scale=0.4]{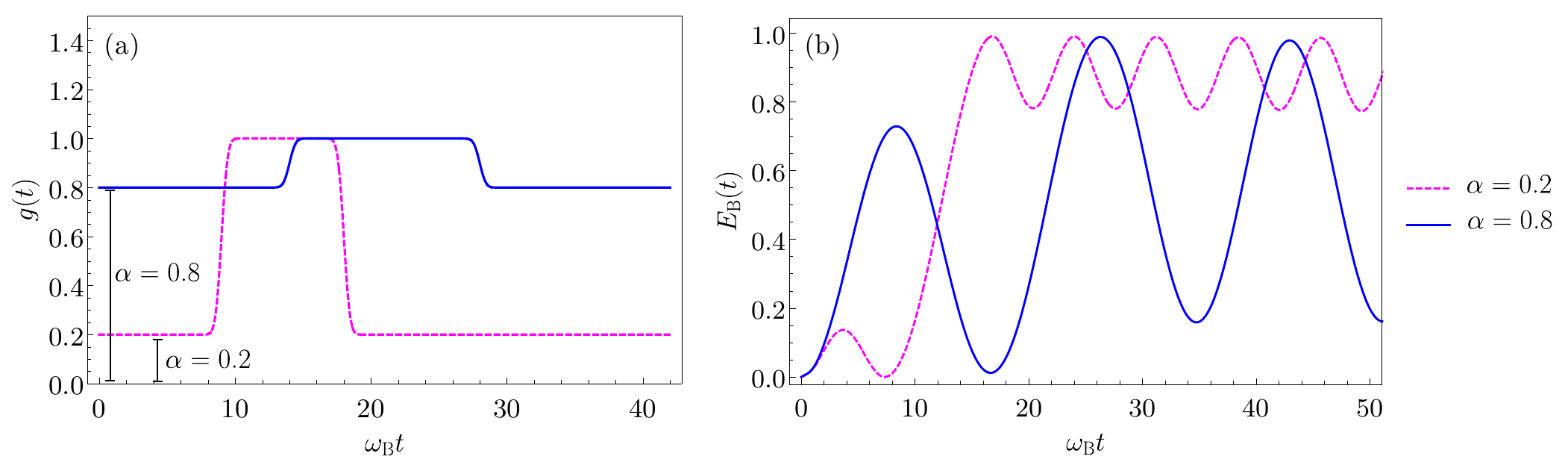} 
   \caption {Panel (a): Behaviour of $g(t)$ as function of $\omega_{\rm B}t$. Panel (b): Behaviour in units of $\omega_{\rm B}$ of $E_{\rm B}(t)$ as function of $\omega_{\rm B}t$. The dashed magenta curves are for $\alpha=0.2$ and $\omega_{\rm B}\tau=14$, while the full blue curves are for $\alpha=0.8$ and $\omega_{\rm B}\tau=9$. All the curves are plotted for $\omega_{\rm B}t_0=0.1$.}
   \label{fig6}
\end{figure}

In the following we are going to rewrite the Hamiltonian for the cavity-mediated model in Eq.~(\ref{HM}). Firstly we need to remap the energy separation of the TLS $\omega_{\rm C}$ and $\omega_{\rm B}$ as 
\begin{equation} \omega_{\rm C}(t)=\alpha \omega_{\rm B}g(t) \quad\quad\quad \omega_{\rm B}(t)=\omega_{\rm B}g(t), \end{equation}
\noindent where $\alpha >0$ is the mismatch parameter previously introduced and $g(t)$ is a smooth function that allows the tuning of the TLSs energy separation of the following form
\begin{equation} g(t)=\frac{(1-\alpha)\bigg[\mathrm{erf}\bigg(\dfrac{t-\tau}{t_0}\bigg)-\mathrm{erf}\bigg(\dfrac{t-2\tau}{t_0}\bigg)\bigg]+2\alpha\mathrm{erf}\bigg(\dfrac{\tau}{2t_0}\bigg)}{2\mathrm{erf}\bigg(\dfrac{\tau}{2t_0}\bigg)}, \end{equation}
where the parameters $\tau$ and $t_0$ are the ones introduced in Eq.~(\ref{f_t}). Examples of the $g(t)$ function are reported in Figure~\ref{fig6} (a), where we can observe that the TLSs are initially off-resonance, then after a time $\tau$ they get on-resonance and when the QB reaches its maximum energy they are again put off-resonance.

Moreover in this kind of protocol the interaction between the charger and the mediator and between the mediator and the QB is time independent, assuming the form
\begin{equation} \tilde H_{\rm int,m}= g[a^\dagger(\sigma_-^{\rm C}+\sigma_-^{\rm B})+a(\sigma_+^{\rm C}+\sigma_+^{\rm B})]. \end{equation}
As a consequence the complete Hamiltonian becomes
\begin{equation}\label{HMt} \tilde H_{\rm m}^{(t)}=\frac{\omega_{\rm C}(t)}{2}\sigma_z^{\rm C}+\frac{\omega_{\rm B}(t)}{2}\sigma_z^{\rm B}+\tilde H_{\rm int,m}. \end{equation}

By diagonalizing and numerically solving the Hamiltonian in Eq.~(\ref{HMt}), using the methods introduced in Ref.~\cite{Crescente22}, we plot the stored energy of the QB in Figure~\ref{fig6} (b). As we can observe, the more the charger and the QB are taken off-resonance (see dashed magenta curve for $\alpha=0.8$) the more stable the storing inside the QB is. This is not true when the mismatch in the TLSs is smaller ($\alpha=0.2$), leading to a continuous transfer of the energy between the charger and the QB and vice versa.

This proves that tuning the coupling between the different parts of the system is a more effective way for a stable storing of the energy inside the QB, compared to changing the mismatch in the energy spacing of the TLSs.

\section{Conclusions}\label{Conclusion}

The analysis of coherent energy transfer processes between a quantum charger and a QB, both modeled as TLSs, has been discussed. Starting from the well known case of direct energy transfer, we have shown how the performances can be improved adding a mediator, in our case a photonic cavity.
Moreover, to consider an experimentally feasible model, we have introduced the possibility of switching on and off the interaction between the parts of the system. This also allows considering the important quantity of the work needed to do such operation.
Considering both on- and off-resonance we have analyzed the stored energy in each part of the system. On-resonance the performances of the direct and cavity-mediated model are similar: both models allow a complete energy transfer, but the photons in the cavity lead to a faster process. In both cases the work in this regime is null, confirming what obtained in Ref.~\cite{Andolina18}.
Different and more interesting are the results obtained off-resonance. Here, the direct energy transfer has very poor performances compared to the cavity-mediated one. In fact by adding more and more photons inside the cavity allows regaining an almost complete energy transfer.
At the end of our work we have also considered a protocol that allows changing the mismatch in TLSs energy level. Here we have proved that this scenario leads to greater instability compared to controlling the coupling constant between each part of the system.

\acknowledgments
The author acknowledge the support of the European Union-NextGenerationEU through the "QUantum Busses for coherent EneRgy Transfer (QUBERT)" project, in the framework of the Curiosity Driven Grant 2021 of the University of Genova.

\end{document}